# CARIOQA: Definition of a Quantum Pathfinder Mission

T. Lévèque*[a], C. Fallet[a], J. Lefebve[a], A. Piquereau[a], A. Gauguet[b], B. Battelier[c], P. Bouyer[c], N. Gaaloul[d], M. Lachmann[d], B. Piest[d], E. Rasel[d], J. Müller[e], C. Schubert[f], Q. Beaufils[g], F. Pereira Dos Santos[g]

[a]Centre National d'Etudes Spatiales, 18 avenue Edouard Belin, 31400 Toulouse, France; [b]LCAR, Université Paul Sabatier, 118 route de Narbonne, 31062 Toulouse, France; [c]LP2N, IOGS, CNRS, Université de Bordeaux, Rue François Mitterrand, 33400 Talence, France; [d]Leibniz Universität Hannover, Institut für Quantenoptik, D-30167 Hannover, Germany; [e]Leibniz Universität Hannover, Institut für Erdmessung, Hannover, Germany; [f]German Aerospace Center (DLR), Institute for Satellite Geodesy and Inertial Sensing, Callinstr. 30b, 30167 Hannover, Germany ; [g]LNE-SYRTE, Observatoire de Paris, Université PSL, CNRS, Sorbonne Université, 61 avenue de l'Observatoire, 75014 Paris, France.

## ABSTRACT

A strong potential gain for space applications is expected from the anticipated performances of inertial sensors based on cold atom interferometry (CAI) that measure the acceleration of freely falling independent atoms by manipulating them with laser light. In this context, CNES and its partners initiated a phase 0 study, called CARIOQA, in order to develop a Quantum Pathfinder Mission unlocking key features of atom interferometry for space and paving the way for future ambitious space missions utilizing this technology. As a cornerstone for the implementation of quantum sensors in space, the CARIOQA phase 0 aimed at defining the Quantum Pathfinder Mission's scenario and associated performance objectives. To comply with these objectives, the payload architecture has been designed to achieve long interrogation time and active rotation compensation on a BEC-based atom interferometer. A study of the satellite architecture, including all the subsystems, has been conducted. Several technical solutions for propulsion and attitude control have been investigated in order to guarantee optimal operating conditions (limitation of micro-vibrations, maximization of measurement time). A preliminary design of the satellite platform was performed.

## 1. INTRODUCTION

### 1.1 Satellite Gravimetry: a unique tool for monitoring climate change

The determination of global mass transport phenomena via gravity field monitoring from satellites is an essential method to tackle the environmental and societal challenge of climate change. Indeed, previous and current space missions (e.g. GRACE, GOCE and GRACE-FO) revolutionized the understanding of mass transport in the Earth system, enabling for the first time the recovery of a global time-varying gravity field. The data obtained in missions so far provides unique and invaluable contributions for understanding climate change processes, such as melting of the glaciers and ice sheets, sea-level rise, regional droughts and flooding, and potentially allow for early warning of such events. The largest error contribution of state-of-the-art missions is linked to the effect of aliasing which results from their incomplete observation geometry. The next generations of gravimetry missions will improve the spatiotemporal sampling of the gravity signal using well-designed satellite constellations, but remain limited in accuracy and resolution. To go one step further, a technological breakthrough involving novel sensors is at hand. It is expected to provide a significant step forward in accuracy. This improvement would pave the way for a global and recurrent remote sensing of essential climate variables related to ocean, cryosphere, atmosphere and land hydrology and monitoring geodynamics phenomena related to Earthquakes and volcanic eruptions.

### 1.2 Space geodesy missions: state-of-the-art and main limitations

Gravity-field missions like GRACE and GRACE-FO provide an invaluable contribution to the understanding of climate change [1,2], closing the sea level rise budget [3] and potentially allowing for early warning of drought or flood events [4]. The main improvement from GRACE to GRACE-FO was the addition of a Laser Ranging Interferometer (LRI) as a technical demonstrator [5] potentially improving the high-frequency performance. However, the performance of the electrostatic accelerometers did not increase significantly. Especially the frequency range below $10^{-3}$ Hz is in part responsible for the characteristic striping pattern found in GRACE gravity field solutions. Other reasons are insufficient background modelling of atmosphere and ocean variability, which will be improved by the next generation of AOD1B product, and the predominantly North-South observation direction. Furthermore, accelerometer performance limits the ability to determine low-degree gravity field coefficients requiring a replacement typically determined by satellite laser ranging [6]. A recent publication on the development of the next generation electrostatic accelerometers shows the overall performance improvement while keeping the decrease in performance in low frequencies [7]. The necessity for calibration of the electrostatic accelerometer remains. Therefore, even if the next generations of gravimetry missions should enable to push further the spatio-temporal sampling of the gravity signal using well-designed satellite constellations, only a technological breakthrough involving novel sensors will enable leap forward in accuracy.

### 1.3 Quantum sensors in space: a technological breakthrough

Classical electrostatic accelerometers used so far in gravimetry missions present increased noise at low frequency and long-term drifts that limit the ability to reconstruct the Earth's gravity field at long wavelength and to accurately model its temporal fluctuations. These limitations have generated a broad interest in disruptive technologies, based on Cold Atom Interferometry (CAI), that measure the acceleration of freely falling independent atoms by manipulating them with laser light. These systems, free from most systematic errors that affect classical systems, provide higher sensitivity at long wavelength [8, 9], drift-free measurements, and higher absolute accuracy, leading to accurate long-term measurements and comparisons. The major interest of bringing quantum accelerometers into space is to extend the interrogation time T and then increase the instrument's sensitivity (scaling as the square of T). On Earth, where the interferometer duration is limited by the fall of the atoms, state-of-the-art CAI has reached an accuracy in the low nano-g range for absolute gravimeters [10, 11] and a differential acceleration sensitivity of order of the nano-g/m for gravity gradiometry [12, 13]. In space, where the quantum superposition in CAIs may be maintained for seconds well beyond what can be achieved on ground, the sensitivity at long wavelength is expected to be several orders of magnitude higher, thus outperforming the best classical devices while contributing to improve measurements of climate change-related mass transport products. In particular, the improved long-term stability of quantum sensors is promising for significantly more accurate understanding of mass transport processes at large scales. Despite its recent progress, this innovative technology suffers from an insufficient of maturity, i.e. a low Technology Readiness Level (TRL), to be ready for operation in space at its best level of performance. Indeed, operating a quantum sensor in space represents a major scientific and technical challenge. First, it is necessary to understand and simulate the performance of these instruments in the context of space gravimetry missions to fully benefit from their performance. Second, the operation of these instruments on a satellite requires specific developments to be adapted to microgravity conditions and a dedicated qualification for the space environment.

In this context, CNES and its partners initiated a phase 0 study, called CARIOQA (Cold Atom Rubidium Interferometer in Orbit for Quantum Accelerometry), in order to develop a Quantum Pathfinder Mission unlocking key features of space atom interferometry and paving the way for future ambitious space missions based on quantum sensors.

## 2. THE CARIOQA QUANTUM PATHFINDER MISSION

### 2.1 Objectives of the Mission

The main objective of the CARIOQA Pathfinder Mission is to demonstrate the operation of a quantum accelerometer in space using a simplified payload. The purpose of this mission is to increase the TRL of the instrument up to 8 by validating the technological building blocks essential to its operation. It will also demonstrate the mastery of the performance of these instruments in space beyond the level that can be achieved on the ground. This is crucial in the context of scientific missions in which these quantum accelerometers are considered. Moreover, even if the primary objective of this mission is the technical validation of quantum sensors in space, secondary scientific objectives relevant for space geodesy are considered. Specifically, the restitution of the low degrees of the Earth's gravity field will be investigated in order to increase the science readiness level to handle Quantum Space Gravimetry data adequately.

Furthermore, the Pathfinder satellite will fly at low altitude in a nadir pointing mode to prepare for Post-Pathfinder gravimetry missions. The satellite platform will be designed to operate the payload at its best level of performances in these conditions (ultra-sensitive attitude control, low noise propulsion, etc.).

The implementation of quantum accelerometers in space for scientific applications requires to bridge a gap of at least two orders of magnitude between the performance of existing ground quantum accelerometers and future space instruments (Table 1). In space, the main gain on the instrument performance is expected from exploiting longer interaction times accessible in microgravity. Nevertheless, recent studies and experiments have shown that reaching these long interaction times in orbit requires implementing innovative technological elements (e.g. ultra-cold atoms, rotation compensation, dedicated design of the satellite platform) that can only be fully tested in space. Indeed, inertial flight environments, including microgravity and related rotation/vibration conditions, are difficult to reproduce on the ground. Therefore, a Quantum Pathfinder Mission is required to bridge the gap between the actual state-of-the-art on-ground performance and the required performance for space applications. This approach agrees with the roadmap for cold atoms in space [14] recently developed and approved by a wide scientific community.

Table 1: Quantum Pathfinder Mission performances with respect to the state-of-the-art

| Instrument | Applications | Sensitivity |
| --- | --- | --- |
| Absolute quantum gravimeter (SYRTE) | Ground/Laboratory | $5 \times 10^{-8}$ m.s$^{-2}$.Hz$^{-1/2}$ |
| Commercial absolute quantum gravimeter (IXBLUE/Muquans) | Ground/Field applications | $5 \times 10^{-7}$ m.s$^{-2}$.Hz$^{-1/2}$ |
| Absolute atom accelerometer/gravimeter in microgravity (ICE/LP2N) | Microgravity | $6 \times 10^{-7}$ m.s$^{-2}$.Hz$^{-1/2}$ |
| **Quantum Pathfinder Mission (CARIOQA)** | **Satellite** | **$1 \times 10^{-10}$ m.s$^{-2}$.Hz$^{-1/2}$*** |
| Quantum Space Gravimetry Missions | Satellite | $1 \times 10^{-12}$ m.s$^{-2}$.Hz$^{-1/2}$ |

*Targeted Quantum projection noise floor*

To ensure the mastery of quantum technology in space, the CARIOQA Pathfinder Mission will implement a quantum accelerometer able to measure the acceleration along the satellite's track. Thus, the CARIOQA Pathfinder Mission has to take up challenges to bring technology beyond the current state-of-the-art by targeting an intermediate level of sensitivity of $1 \times 10^{-10}$ m.s$^{-2}$.Hz$^{-1/2}$, corresponding to the Quantum projection noise floor (i.e. the fundamental limit for the intrinsic noise of the instrument). Reaching this performance in space will outperform ground sensors and pave the way to Post-Pathfinder scientific missions.

### 2.2 Mission characteristics

The CARIOQA mission will consist in the implementation of a single-axis atom interferometer onboard a dedicated satellite platform. The satellite will fly at an altitude of 600 km on a sun synchronous orbit with no orbit maintenance system to simplify the satellite architecture. The satellite will be operated in a nadir pointing mode in order to be representative of gravity missions conditions. The payload will be accommodated in such a way that the atoms are manipulated at the center of mass of the satellite in order to avoid rotation related bias on the measurement. The measurement axis of the instrument (i.e. the atom accelerometer) will be oriented along the velocity axis of the satellite.

As a Quantum Pathfinder Mission, CARIOQA is not designed for measuring the Earth's gravity field. Indeed, for this purpose, two accelerometers are generally required to form a differential measurement, like in the GOCE and GRACE missions. Therefore, monitoring the acceleration along the velocity axis of the satellite will provide a measurement of non-gravitational forces acting on the satellite (i.e. drag force and sun radiation pressure). This typical input signal will be used for characterizing the instrument, especially its dynamics. The onboard characterization of the instrument will consist in the implementation of several measurement modes enabling to determine each source of noise independently. In particular, the characterization process will include dedicated AOCS manoeuvers for the characterization of rotation bias on the instrument.

## 3. INSTRUMENT CHARACTERISTICS

Developing out a Quantum Pathfinder Mission to test a quantum accelerometer in space places two major constraints on this instrument. First, it must be operational on board a satellite platform, and therefore, in a microgravity space environment. Second, such a mission would aim to increase the instrument's performance in space, which implies the exploitation of long interaction times while controlling the interferometer phase.

## 3.1 Instrument description and trade off

Considering the CARIOQA mission objectives, several payload options have been investigated. In particular, different sorts of atomic source technologies have been compared considering their potential performances with respect to space atom interferometry, their compatibility with space application, and their TRL. As a result, the utilisation of a BEC atomic source has been chosen as it enables a higher mitigation of rotation effects and then enables to demonstrate significantly higher instrument performances. For producing the BEC, the choice of atom chip technology has been realized as it presents the highest TRL for space application [15, 16].

The CARIOQA Pathfinder Mission payload will then consist in a one-axis quantum accelerometer based on ultra-cold Rubidium atoms manipulated by Raman transitions. Rubidium atoms are first prepared from a vapor to form an ultra-cold sample (BEC), using an atom-chip. The atomic cloud is then moved away from the surface of the chip and collimated via magnetic lensing. Afterwards, the free-falling Rubidium cloud successively interacts three times with a unique pair of retro-reflected Raman beams, acting on matter waves as beam splitters or mirrors. This creates an interferometer of 2T total interaction time in a so-called double-diffraction configuration [17]. This interferometer scheme, depicted on Figure 1, is particularly adapted to space instruments. Indeed, for inertial sensors in microgravity environments, the Doppler Effect cannot be used to select one or the other of the two effective Raman transitions possible in the most common retro-reflected configuration, which is needed for accuracy. The atomic phase shift is obtained from the measurement of populations in each output port of the interferometer, using a fluorescence or absorption imaging technique. To detect the atoms at the output of the interferometer, two complementary methods are implemented including a CCD camera and a photodiode. The CCD camera on the primary imaging axis is oriented perpendicular to atom interferometer axis and simultaneously enable spatially resolved detection of both output ports.

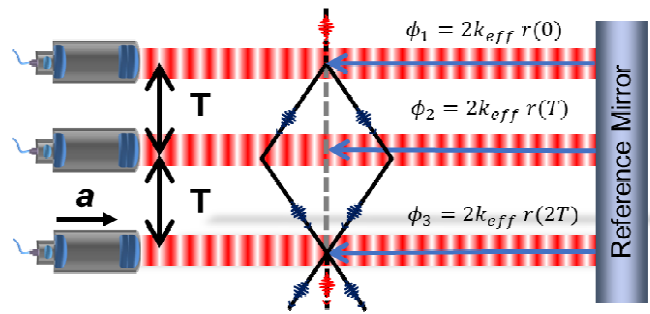

Figure 1: Scheme of a three pulses atom interferometer using Raman lasers in a retro-reflected configuration. The total interaction time is 2T. The interferometer is sensitive to the acceleration of the cold atoms with respect to the retro-reflection mirror in the direction of the Raman lasers.

## 3.2 System Architecture:

The instrument is composed of four major subsystems depicted on Figure 2. The Physics Package includes the vacuum system, collimators and coils among others. This is the instrument's heart where the atoms will be cooled, interrogated and detected. The Laser System provides the different laser beams needed to cool, interfere and detect the atoms. In this purpose, it shall provide laser light stable in both frequency and power while ensuring an accurate control of these parameters. The Microwave Source produces the radio and microwaves frequencies needed to adjust the laser frequencies, and the master clock of the instrument. The Instrument Control and Power Unit (ICPU) provides the power and controls the instrument by driving the various measurement sequences.

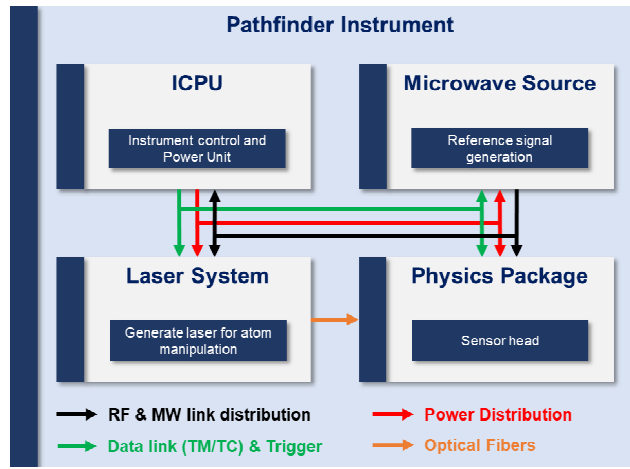

Figure 2: Block diagram of the Pathfinder instrument showing its four subsystems (ICPU, Laser System, Physics Package, and Microwave Source) and their interactions.

For space applications, contrary to existing ground instruments, the design of a quantum accelerometer has to consider the absence of relative displacement of the atoms inside the vacuum system of the Physics Package [18]. Consequently, the complete experimental sequence is achieved at a single location inside the vacuum chamber. This leads to a substantial advantage in terms of compactness. Nevertheless, this implies that the interrogation zone is not spatially separated from the atom source, potentially leading to spurious magnetic fields, parasitic light at the location of the atoms, and higher atom losses due to background gas. Benefitting from the fall of the atoms, the different phases of atom manipulation are well mastered separately on the ground. On the contrary, their combination at the same place requires an optimized design dedicated to microgravity. Another specificity in microgravity is the double diffraction regime (in the usual Raman laser pairs retroreflected by a reference mirror) because of the absence of Doppler effect, since the atoms remain at rest after preparation. Practical experimental techniques need to be adapted such as the control of the atomic phase, which can no longer be adjusted simply by controlling the laser phase difference.

### 3.3 Performance analysis:

Achieving long interrogation time implies using ultra-cold gases at very low temperatures (<100 pK) to maintain the atom interferometer's contrast and signal-to-noise [19]. However, this comes at the expense of a low atom flux, thus limiting the performance gain using ultra-cold atoms. Indeed, state-of-the-art quantum gravimeters/accelerometers are currently based on thermal clouds as the atomic sources at intermediate temperature regime (~µK). Improving the flux of the ultra-cold atom source is then a major challenge. Beyond enhancing the instrument sensitivity of several orders of magnitude, the goal is to demonstrate a corresponding accuracy improvement. Moreover, it leads to a series of technical challenges in terms of high-quality optics (e.g., wave front flatness) [20], laser beam intensity profiles, control of the position and the velocity of the atom gas [21] and magnetic environment beyond the usual characteristics of ground-based atom interferometers. For applications in geodesy, the satellite platform needs to operate in a nadir pointing mode. Rotation of the satellite in this mode leads to a rotation of the reference mirror that needs to be compensated to avoid contrast loss and spurious phase shifts due to undesirable inertial terms (e.g. Coriolis, centrifugal force). Implementation of an accurate rotating mirror mount [11] is necessary to tackle this difficulty. Last but not least, the instrument design needs to be adapted to the requirements of a satellite mission. This includes a reduction of the volume, mass and power consumption of the instrument and a rugged design with respect to environmental conditions of space flight such as vibrations, thermal vacuum conditions, and radiations. In addition, the instruments' reliability and autonomy need to be addressed, as operation on a satellite does not allow for maintenance or re-adjustment of the apparatus. Due to the complexity of the instrument, this adaptation constitutes a technological challenge by itself.

The instrument shall be designed to explore acceleration sensitivity lower than $10^{-9}$ m.s$^{-2}$.Hz$^{-1/2}$. In order to do so, the main characteristics of the instrument will be chosen to target a quantum projection noise floor corresponding to a sensitivity below $10^{-10}$ m.s$^{-2}$.Hz$^{-1/2}$. This noise floor can be calculated by considering the main parameters of the atom interferometer presented in Table 2. These parameters lead to a shot-noise limited sensitivity of $3 \times 10^{-10}$ m.s$^{-2}$.Hz$^{-1/2}$

(conservative value) which could be improved to $8 \times 10^{-11}$ m.s$^{-2}$.Hz$^{-1/2}$ (enhanced value) by increasing the atom number in the BEC up to $10^6$. Production of BECs with $10^5$ Rubidium atoms in 1 s [22] and a total kinetic energy in 3D of $(3/2) k_B$ 38 pK [23] were experimentally demonstrated with setups using atom chips.

Table 2: Specification of the parameters assumed for the performance analysis

| Parameter | Symbol | Value | Unit |
| --- | --- | --- | --- |
| Interaction time | 2T | 2 | s |
| Contrast | C | ~0.8 | - |
| Preparation time | $t_{prep}$ | 1.5 | s |
| Cycle duration | $t_c$ | 3.5 | s |
| Detected atom number | $N_{det}$ | $10^5$ | - |
| Effective temperature | Θ | $100 \times 10^{-12}$ | K |
| Accelerometer sensitivity (conservative) | $\sigma_a$ | $3 \times 10^{-10}$ | m.s$^{-2}$.Hz$^{-1/2}$ |
| Accelerometer sensitivity (enhanced) | $\sigma_a$ | $8 \times 10^{-11}$ | m.s$^{-2}$.Hz$^{-1/2}$ |

# 4. SATELLITE CHARACTERISTICS

The CARIOQA Pathfinder Mission shall enable to fully characterize the payload and its capabilities in a suitable environment. For this reason, the mission will be implemented on board a dedicated satellite platform. This will allow the satellite design and mission parameters to be optimized to best meet the payload requirements. In particular, the instrument will be accommodated at the center of mass (COM) of the satellite and the AOCS system will be chosen to optimize the rejection of rotations and the performance of the instrument. In addition, the choice of a dedicated mission will enable to choose an orbit allowing the best characterization of the instrument and to consider maneuvers dedicated to its calibration. A preliminary design of the satellite platform has been carried out, fulfilling the mission requirements while keeping its costs as low as possible.

## 4.1 Main requirements of the satellite platform

In order to prepare future EO Mission, the CARIOQA Pathfinder instrument shall be designed to fit with a medium class satellite platform. The specificities of the technological pathfinder mission lead to particular constraints on the satellite.

First, given the extreme sensitivity of atom accelerometers, dynamic disturbances (vibrations, cracking, thermos-elastic, …) shall be avoided onboard the satellite plateform. In order to comply with payload performance requirements, the integrated effect of dynamic disturbances onto the instrument measurement (taking into account its sensitivity function) shall remain below $10^{-10}$ m.s$^{-2}$. This constraint mainly affects the attitude control subsystem. Indeed, standard AOCS actuators (reaction wheels) are known for generating microvibrations. Thus, a specific analysis of their impact on the accelerometer measurement showed that reaction wheels are not compatible with the mission requirements. Therefore, thrusters must be used as AOCS actuators.

Second, the main requirement concerning the accommodation of the instrument onboard the platform is related to its position with respect to the COM of the satellite. Indeed, in order to minimize the effect of the nadir rotation onto the systematic effects of the instrument, the atomic cloud shall be manipulated at the COM of the satellite. To guarantee the overall performance of the instrument, the distance between the atomic cloud and the COM shall remain under 1 mm while being monitored at +/- 0.1 mm. In addition, the residual velocity of the atomic cloud with respect to the satellite shall remain below +/- 0.01 mm/s. These requirements have a significant impact on the satellite architecture and instrument accommodation, resulting in a strong interpenetration of the payload and the platform. It also constrains the type of propulsion that can be used. Indeed, a liquid propulsion (hydrazine type) implies a change in the COM as the tank is emptied. In addition, liquid sloshing can induce accelerations that would disturb the measurement of the accelerometer. As a result, electric propulsion has been chosen as AOCS actuators.

Finally, in order to ensure the efficiency of the rotation compensation system inside the instrument, the nadir orientation shall be kept with stringent pointing performance. The pointing error shall remain under 0.5° on each axis. The angular speed errors (difference between commanded and real angular speed) shall remain respectively under 5. $10^{-6}$ rad/s on the orbital direction and 5. $10^{-5}$ rad/s on the two other axis. The angular speed knowledge (difference between ground estimated angular speed and real angular speed) shall be better than $5 \cdot 10^{-6}$ rad/s on each axis. These requirements have a strong impact on the attitude estimation.

## 4.2 Satellite preliminary design

Due to the constraint of placing the heart of the instrument at the COM of the satellite, off-the-shelf platform cannot be considered. However, the Myriade Evolutions structure, which has a compatible ring interface with the launcher adapters, is used as a reference for this platform. The preliminary design of the satellite, depicted on Figure 3, is based on off-the-shelf equipment (Myriade Evolutions) in order to reduce the cost of the mission.

The envisioned platform provides all the necessary housekeeping functions to achieve the mission goals: a payload support, electrical power, thermal control, command, data handling and storage, attitude and orbit control. Two solar panels are coupled with batteries in order to provide the energy to the platform. The data handling architecture is based on a central computer (OBC). The housekeeping TM/TC (telemetry/telecommand) and the payload telemetry is realised in S band. For the precise orbit determination, a GNSS receiver is also integrated to the platform. A preliminary study of mass, consumption and volume was conducted to ensure the feasibility of the mission.

In order to obtain an optimal mass distribution, particular attention is paid to the accommodation of the thrusters so as to compensate for the mass of metal consumed in order to minimize the displacement of the COM along the mission. Nevertheless, center of mass trims (CMT) will be implemented to allow the position of the center of mass to be readjusted during the satellite lifetime.

For thermal constraints, the complete payload is accommodated on the panel which is not exposed to the Earth's albedo (i.e. the upper panel). The Physics Package, where the atoms are manipulated, is placed inside the structure, at the COM of the satellite, while other payload subsystems are accommodated outside the structure.

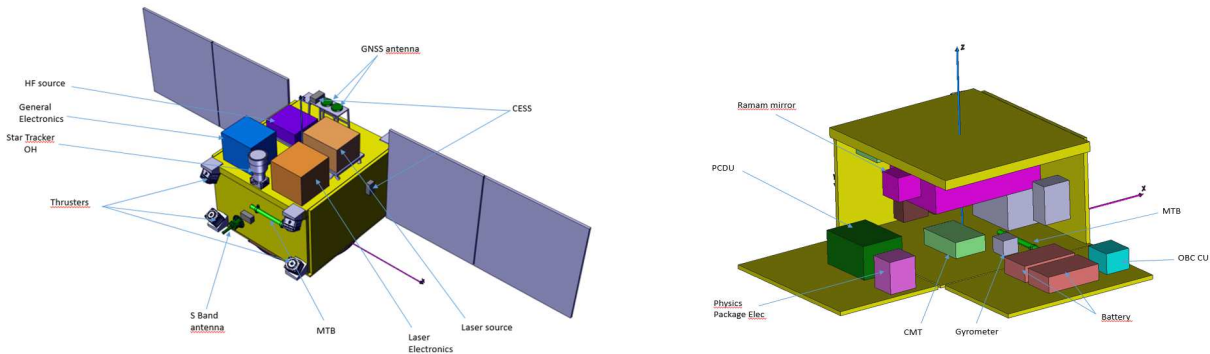

Figure 3: Preliminary design of the CARIOQA satellite platform.

The AOCS has been developed to comply with the performance requirements of the payload. In particular, the most stringent performance requirement is on the angular speed error, which has to remain below $5.10^{-6}$ rad/s. For that, a preliminary analysis has shown the necessity of a gyro-stellar attitude estimation to reduce the star tracker (SST) noise and to get a better attitude estimation. A simplified simulator dedicated to this study has been developed. This simulation uses a gyro-stellar filter that mixes the measurement from the SST with the one from the gyroscope. The AOCS equipment considered in this model are the following ones:

- 8 electric thrusters (Maximum thrust: 350 μN, ISP: > 3000s, Mass dry/wet: 680/900g).
- 1 star tracker (Noise 1σ XY/Z: 12 μrad / 96 μrad)
- 1 gyroscope (ARW: 0.005 deg/sqrt(h), Constant bias: 0.3 deg/h, Bias stability: 0.02 deg/h over one hour, Scale factor: 500ppm).

The result of the simulation shows that the standard deviation of the angular error is $2 \cdot 10^{-6}$ rad/s and the maximum values is $8 \cdot 10^{-6}$ rad/s. The hardware setup that has been selected is then compatible with the requirements of $5 \cdot 10^{-6}$ rad/s even if further improvements and optimizations of the control laws are needed.

# CONCLUSION

The CARIOQA Phase 0 study demonstrated the feasibility of a Quantum Pathfinder Mission dedicated to quantum sensors. First, the main specifications of this mission (objectives, definition of the instrument, main performances, mission parameters) have been determined and justified by analysis. Thus, a mission scenario was established based on these specifications. A preliminary definition of the payload and the satellite platform has been performed. These studies focused in particular on the definition of critical parameters to ensure the performance of the instrument and its compatibility with the platform (AOCS performance). Then, the various constraints resulting from this definition enabled to identify the optimal orbit for the mission.